# Exciton magnetotransport in two-dimensional systems: weak-localization effects

P. I. Arseev

*P. N. Lebedev Institute of Physics, Russian Academy of Sciences, 117924 Moscow, Russia*

A. B. Dzyubenko*[)]

*Institute of General Physics, Russian Academy of Sciences, 117942 Moscow, Russia*

The paper considers the effect of a magnetic field $B$ on the transport of neutral composite particles, namely excitons, in weakly disordered two-dimensional (2D) systems. In the case of classical transport (when the interference of different paths is neglected), the magnetic field suppresses exciton transport, and the static diffusion constant $D(B)$ monotonically drops with $B$. When quantum-mechanical corrections due to weak localization are taken into account, $D(B)$ becomes a nonmonotonic function of $B$. In weak magnetic fields, where the magnetic length is much larger than the exciton Bohr radius, $\ell_B = (\hbar c/eB)^{1/2} \gg a_B = \varepsilon \hbar^2/\mu e^2$, a positive magnetodiffusion effect is predicted, i.e., the exciton mobility should increase with $B$. © *1998 American Institute of Physics.* [S1063-7761(98)02407-X]

## 1. INTRODUCTION

In two-dimensional (2D) systems, all states are localized, no matter how weak the disorder is.[1–3] This phenomenon is universal for all processes of wave propagation and is associated with the constructive interference of paths subjected to the time-reversal operation. For particles with nonzero mass, this is a quantum-mechanical effect, which cannot be described in terms of classical mechanics. The quantum statistics of particles in this case does not play a crucial role (see, for example, works on the weak localization of phonons[4] and light[5]). While localization has been thoroughly investigated on the base of the one-particle approach, many questions concerning the interplay between localization and Coulomb effects remain unanswered. The variety of physical situations requires the application of different techniques suitable for describing the respective class of phenomena. For instance, it was predicted by the weak-localization theory that the electron-electron interaction should weaken the interference effects and lead to a higher conductivity (see, e.g., the review by Lee and Ramakrishnan[3]). A numerical calculation for two interacting electrons in a random potential has also predicted a correlated-propagation length larger than the localization length of an isolated particle.[6] The issue discussed in this paper, namely the propagation of an exciton, which consists of an electron and a hole interacting with one another, in a magnetic field and in a random potential is also one of the aspects of the general problem. The weak localization of excitons in the absence of a magnetic field was investigated previously.[7]

Introduction of a magnetic field $B$ generates new features in the physical picture of the weak localization of electrons. Formally, a magnetic field $B$ breaks time-reversal symmetry. The physical consequence is negative magnetoresistance in electron systems.[8,9] This effect is caused by the fact that charged particles acquire different phase shifts in magnetic fields when they travel along closed paths in opposite directions.[10] As a result, the field $B$ breaks the constructive interference between time-reversed paths and thereby suppresses the weak localization of electrons. If we take into account the electron spin, four different channels for interference between two electronic waves are possible: one of them is singlet ($S=0$), and three are triplet ($S=1$, $S_z = \pm 1, 0$). The interference in the triplet (singlet) channels gives a positive (negative) contribution to the conductivity.[9,11] Fast spin-flip processes can change the relation between the contributions of the singlet and triplet channels, thus resulting in either negative or positive magnetoresistance. Various mechanisms of spin-orbit coupling that are important for electrons in quasi-two-dimensional semiconductor quantum wells and heterojunctions were taken into account in Ref. 12. Note also that in systems with strongly localized electron states (the hopping conductivity regime) magnetically induced changes in the phase relations between different transition amplitudes can lead to either negative or positive magnetoresistance.[13,14]

An important question in the case of excitons, which are composite and, as a whole, electrically neutral particles, is whether the time-reversal symmetry for an $e$–$h$ pair is broken by a magnetic field $B$. One may assume that the $t \to -t$ symmetry for a pair should be broken by magnetic field since it is broken for an electron or hole taken separately, and this is true in a general case. There is, however, an exceptional case. Consider the Hamiltonian

$$H = \frac{1}{2m_e}\left(-i\hbar\nabla_e + \frac{e}{c}\mathbf{A}_e\right)^2 + \frac{1}{2m_h}\left(-i\hbar\nabla_h - \frac{e}{c}\mathbf{A}_h\right)^2 + U_{eh}(\mathbf{r}_e - \mathbf{r}_h) + V_e(\mathbf{r}_e) + V_h(\mathbf{r}_h), \qquad (1)$$

which describes the motion of an $e$–$h$ pair in a uniform magnetic field $B$ and external (random) potentials $V_e$ and $V_h$. When the particle masses are equal, $m_e = m_h$, and the scattering potentials are identical, $V_e \equiv V_h$, the $e$ and $h$ components transform into one another after time reversal.[1)] In this case Hamiltonian (1) commutes with the time-reversal operator:

$$[H, \hat{T}] = 0. \qquad (2)$$

This means that the $t \to -t$ symmetry is not broken, and 2D excitons should remain localized even in the presence of a magnetic field.

In the general case, one should analyze how a magnetic field $B$ suppresses the weak localization of excitons, which are electroneutral as a whole, and how their internal structure manifests itself. In a magnetic field, the center-of-mass motion and relative motion of an $e$–$h$ pair are coupled. Therefore, the scattering of an exciton as a whole is affected by the magnetic field $B$ and the internal $e$–$h$ interaction.

Recently, the transport of quasi-two-dimensional excitons in quantum wells under a magnetic field has attracted a lot of attention on the part of experimentalists (see Refs. 16–18 and references therein). Butov *et al.*[17] reported intriguing low-temperature anomalies in exciton magnetotransport. In particular, they found that the exciton diffusion constant $D$ is a nonmonotonic function of $B$ and increases considerably in the range of intermediate fields $B \simeq 6$ T. This fact was interpreted as evidence in favor of Bose–Einstein condensation and a manifestation of the superfluidity of excitons. It seems interesting to check whether the localization effects of excitons can give rise to such features of the $D(B)$ curve in the normal phase. In this paper we investigate theoretically the magnetotransport of 2D excitons in the presence of weak disorder in the limiting case when the magnetic length is much larger than the Bohr radius of the exciton, $\ell_B = (\hbar c/eB)^{1/2} \gg a_B = \varepsilon \hbar^2/\mu e^2$. Fields that satisfy this condition will be dubbed weak. Taking these results together with those for the cases of classical[19] and quantum transport[20,21] in the opposite limit, $\ell_B \ll a_B$, we shall suggest an approximate form of the diffusion constant $D$ as a function of $B$ at all fields, including the intermediate range, where $\ell_B \sim a_B$. A brief account of some results of this paper was reported previously.[21]

## 2. EXCITON TRANSPORT IN A MAGNETIC FIELD $B$

### 2.1. Problem statement

In the weak-localization regime, the interaction with an isolated defect does not give rise to a bound state, and localization is possible only at large distances due to the interference of scattered waves. This localization regime takes place in the case of weak scattering, in which

$$\gamma_0(p) \ll \epsilon(p), \qquad (3)$$

where $\gamma_0$ is the damping coefficient (the reciprocal of the momentum relaxation time) of an exciton with energy $\epsilon(p)$. The scatterers in this case are the random potentials $V_e(\mathbf{r})$ and $V_h(\mathbf{r})$ in Eq. (1), which act on the electron and hole separately. They can be, e.g., potentials generated by charged impurities, effective potentials due to irregularities on quantum-well interfaces, etc. At low temperatures, the dominant scattering mechanism in quantum wells is that due to irregularities on the interfaces ("surface roughness"), and our attention will be mainly focused on this mechanism. In the case of 2D excitons in a quantum well of width $d$ in the presence of interface irregularities with a characteristic amplitude $\Delta$ and a correlation length $\Lambda$ (see Ref. 19 and references therein), there is a characteristic exciton momentum defined as follows:[2)]

$$p_{\min} \sim \frac{1}{a_B}\left(\frac{\Delta \Lambda a_B}{d^3}\right), \quad \ell_B \gg a_B. \qquad (4)$$

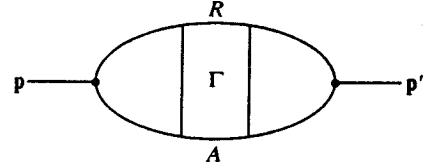

FIG. 1. Diagrammatic representation of Eq. (5): $\Gamma$ is the irreducible vertex corresponding to scattering by a random potential; the lines labeled by $A$ and $R$ represent the advanced and retarded propagators $G^A$ and $G^R$ of excitons averaged with respect to disorder in a magnetic field. The current vertices in the diagram correspond to the exciton center-of-mass velocity $V(\mathbf{p})$.

For long-wavelength excitons with momenta $p < p_{\min}$ we have $\gamma_0 \gtrsim \epsilon$, and such excitons are strongly localized. An analysis of the strong localization of composite particles in a magnetic field is beyond the scope of this paper. However, if the parameter $\Delta \Lambda a_B/d^3 \ll 1$ is sufficiently small, the range of strong localization of excitons is narrow compared with the characteristic momentum $a_B^{-1}$ in our problem ($\ell_B^{-1}$ in high magnetic fields), and the theory developed in this paper has a region of applicability.

It is essential that the scattering of two-particle $e$–$h$ states can be described diagrammatically in terms of effective one-particle (exciton) scattering (this approach is justified in Appendix A). This approximation allows us to treat excitons at low densities as Bose-particles. Their internal structure manifests itself in changes in the effective scattering potential $V_{\mathbf{p},\mathbf{p}'}$ and the dispersion law $\epsilon(p)$ due to the magnetic field $B$ (see Sec. 2.2). The potentials $V_e(\mathbf{r})$ and $V_h(\mathbf{r})$ may be uncorrelated, for example, when the particles $e$ and $h$ are spatially separated,[19] or fully correlated, e.g., when both particles are in the same spatial domain. We assume that the distribution of random fields is Gaussian and use standard diagram techniques[23] with two-particle [retarded ($R$) and advanced ($A$)] excitonic propagators in a magnetic field averaged with respect to disorder: $G_\omega^{R(A)}(p) = [\omega - \epsilon(p) \pm i\gamma_0(p)]^{-1}$ (see, e.g., Ref. 20).

For the case of elastic scattering, one can introduce a diffusion constant $D(\omega, \epsilon)$ for excitons with a given energy $\epsilon$ at a frequency $\omega$, which can be derived from the expression for the generalized "conductivity" (Fig. 1) $\sigma(\omega, \epsilon) = D(\omega, \epsilon) \mathcal{N}(\epsilon)$:

$$\sigma(\omega, \epsilon) = \frac{1}{2\pi} \int d\mathbf{p} \int d\mathbf{p}' \langle\langle V_x(\mathbf{p}) G^R(\mathbf{p}, \mathbf{p}', \epsilon + \omega) \\ \times G^A(\mathbf{p}', \mathbf{p}, \epsilon) V_x(\mathbf{p}') \rangle\rangle, \qquad (5)$$

where $\mathbf{V}(\mathbf{p})$ is the velocity of the exciton's center of mass, $\mathcal{N}(\epsilon)$ is the exciton density of states, and $\langle\langle \ldots \rangle\rangle$ denotes averaging with respect to disorder. Note that the diffusion constant $D(\omega,\epsilon)$ is a parameter included in the "diffusion" pole of the exciton "density-density" correlation function and, therefore, determines the propagation characteristics of particles with a given energy $\epsilon$ in the long-wavelength limit. The localization of quantum states with energy $\epsilon$ means that the diffusion constant as a function of frequency, $D(\omega)$, tends to zero in the static limit, $\omega \to 0$. If inelastic scattering it taken into account, $D(\omega)$ turns out to be finite (see, for example, the review by Lee and Ramakrishnan[3]). The static limit $D(\epsilon)=D(\epsilon,\omega=0)$ is determined by the time $\tau_\phi$ of the loss of phase coherence (dephasing). The *total* static diffusion constant $D=D(T)$ corresponding to fluctuations in the density of excitons with all allowed energies can be obtained from the microscopic values of $D(\epsilon)$ using the generalized Einstein relation:

$$D = \frac{\int d\epsilon \mathcal{N}(\epsilon)[-\partial f/\partial \epsilon]D(\epsilon)}{\int d\epsilon \mathcal{N}(\epsilon)[-\partial f/\partial \epsilon]}, \tag{6}$$

where $f=f(\mu_X,T)$ is the distribution function and $\mu_X$ is the chemical potential of the excitons.

### 2.2. Effective scattering potential

The Hamiltonian $H_0$ of relative motion of an $e-h$ pair with center-of-mass momentum $\hbar \mathbf{p}$ (where $\mathbf{p}$ is the wave vector) in a perpendicular magnetic field $B$ has the form[24,25]

$$H_0 = -\frac{\hbar^2}{2\mu}\nabla_\mathbf{r}^2 - \frac{i\hbar eB}{2c}\left(\frac{1}{m_h}-\frac{1}{m_e}\right)(\mathbf{r}\times\nabla_\mathbf{r})_z + \frac{e^2 B^2}{8\mu c^2}r^2$$
$$+ \frac{e\hbar}{Mc}\mathbf{B}\cdot(\mathbf{r}\times\mathbf{p}) - \frac{e^2}{\varepsilon|\mathbf{r}|}, \tag{7}$$

where $\mathbf{r}=\mathbf{r}_e-\mathbf{r}_h$ is the relative $e-h$ coordinate, and $\mu^{-1}=m_e^{-1}+m_h^{-1}$. In writing this expression, we have utilized the existence of an exact integral of motion, namely the magnetic center-of-mass momentum[24] defined by the operator

$$\hbar\hat{\mathbf{p}} = -i\hbar\nabla_\mathbf{R} - \frac{e}{c}\mathbf{A}(\mathbf{r}),$$

where $\mathbf{R}=(m_e\mathbf{r}_e+m_h\mathbf{r}_h)/M$ is the center-of-mass coordinate, $M=m_e+m_h$, and the vector potential is taken in the symmetrical gauge $\mathbf{A}=\mathbf{B}\times\mathbf{r}/2$. The exciton wave function in a magnetic field $B$ has the form

$$\Psi_\mathbf{p}(\mathbf{R},\mathbf{r}) = \exp\left\{i\mathbf{R}\left[\mathbf{p}+\frac{e}{c}\mathbf{A}(\mathbf{r})\right]\right\}\Phi_\mathbf{p}(\mathbf{r}). \tag{8}$$

An important point is that the wave function $\Phi_\mathbf{p}$ of relative motion of an $e-h$ pair depends on the center-of-mass momentum $\mathbf{p}$,[24] i.e., the relative motion and the center-of-mass motion are coupled. The scattering matrix elements between the exciton states with the center-of-mass momenta $\mathbf{p}$ and $\mathbf{p}'$ in an external potential $\hat{V}=V_e(\mathbf{r}_e)+V_h(\mathbf{r}_h)$ have the form (see Appendix A)

$$V_{\mathbf{p},\mathbf{p}'} = \langle \Psi_\mathbf{p}|\hat{V}|\Psi_{\mathbf{p}'}\rangle. \tag{9}$$

In this work we use an approximation that ignores transitions to excited states of internal motion.[19,20] In the weak-field limit, $\ell_B \gg a_B$, the problem can be treated analytically.[3] We calculate the ground-state wave function $\Phi_\mathbf{p}(\mathbf{r})$ in a magnetic field using perturbation theory with respect to terms containing the magnetic field in the Hamiltonian (7) of the relative motion of an $e-h$ pair, and then we obtain the scattering matrix elements $V_{\mathbf{p},\mathbf{p}'}$. They can be expressed as

$$V_{\mathbf{p},\mathbf{p}'} = F^e_{\mathbf{p},\mathbf{p}'}\tilde{V}_e(\Delta\mathbf{p}) + F^h_{\mathbf{p},\mathbf{p}'}\tilde{V}_h(\Delta\mathbf{p}), \tag{10}$$

where $\tilde{V}_j(\mathbf{p})$ are two-dimensional Fourier transforms of the potentials $V_j(\mathbf{r})$ ($j=e,h$), $\Delta\mathbf{p}=\mathbf{p}'-\mathbf{p}$ is the momentum transfer,

$$F^{e(h)}_{\mathbf{p},\mathbf{p}'} = \int d\mathbf{r}\Phi_\mathbf{p}^*(\mathbf{r})\Phi_{\mathbf{p}'}(\mathbf{r})\exp\left\{\pm i\frac{m_{h(e)}}{M}(\mathbf{p}'-\mathbf{p})\mathbf{r}\right\} \tag{11}$$

are the form factors related to the wave function of the internal motion of the exciton. In the weak-field limit, we must calculate the wave functions up to the second order in $B$ and then substitute them into Eqs. (11) and (10) (see Appendix B). Note that the exponential function in Eq. (11) can be expanded in powers of its argument when $p$, $p' \ll a_B^{-1}$, and only terms of the lowest orders need be included. The limitation on the momenta is essential if we do not take into consideration transitions to excited states. Indeed, if $p$, $p' \sim a_B^{-1}$, the exciton kinetic energy is sufficient for transitions to excited states of internal motion, which are excluded from our analysis.

Taking the essential terms of up to the second order in $B$ and the lowest orders in $pa_B$ of interest to us, we obtain

$$V_{\mathbf{p},\mathbf{p}'} = \bar{V}_e(\Delta\mathbf{p})\left[1+\beta_e(\Delta\mathbf{p})^2 a_B^2\left(\frac{a_B}{\ell_B}\right)^4 - i\alpha_e[\mathbf{pp}']_z a_B^2 \right.$$
$$\left. \times\left(\frac{a_B}{\ell_B}\right)^2\right] + \bar{V}_h(\Delta\mathbf{p})\left[1+\beta_h(\Delta\mathbf{p})^2 a_B^2\left(\frac{a_B}{\ell_B}\right)^4 \right.$$
$$\left. + i\alpha_h[\mathbf{pp}']_z a_B^2\left(\frac{a_B}{\ell_B}\right)^2\right]. \tag{12}$$

Here $\bar{V}_i(\Delta\mathbf{p}) = \tilde{V}_i(\Delta\mathbf{p})F^i_{\mathbf{p},\mathbf{p}'}(B=0)$,

$$F^{e(h)}_{\mathbf{p},\mathbf{p}'}(B=0) = \left\{1+\frac{1}{16}\left[\frac{m_{h(e)}(\mathbf{p}-\mathbf{p}')a_B}{M}\right]^2\right\}^{-3/2}$$

is the form factor corresponding to the ground-state wave function of the 2D-exciton at $B=0$.

An important point is that time-reversal symmetry is broken for this effective scattering potential:

$$V_{\mathbf{p},\mathbf{p}_1} \neq V_{-\mathbf{p}_1,-\mathbf{p}}, \tag{13}$$

the only exception being the case of $V_e=V_h$ and $m_e=m_h$ [see Eq. (2)]. Equation (12) contains the dimensionless constants

$$\beta_{e(h)} = -\frac{\mu^2}{2M^2}\frac{\hbar^4}{\mu^2 a_B^6}\sum_n{}'\frac{|\langle 0|x|n\rangle|^2}{(\epsilon_0-\epsilon_n)^2}$$
$$+ \frac{m_{h(e)}^2}{8M^2}\frac{\hbar^2}{\mu a_B^6}\sum_n{}'\frac{|\langle 0|r^2|n\rangle|^2}{\epsilon_n-\epsilon_0} \tag{14}$$

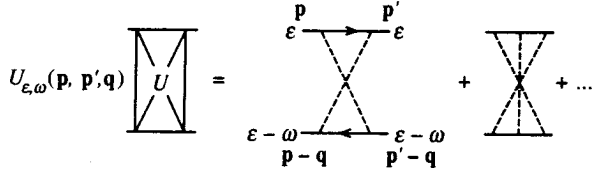

FIG. 2. Sum of maximally crossed diagrams $U_{\epsilon,\omega}(\mathbf{p},\mathbf{p}',\mathbf{q})$. The upper (lower) line corresponds to the retarded (advanced) propagator $G^{R(A)}$ of an exciton averaged with respect to disorder.

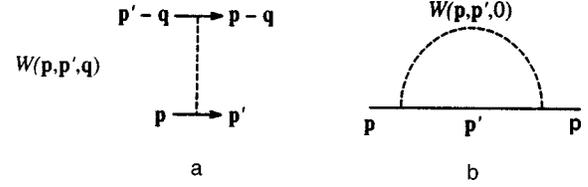

FIG. 3. (a) Simplest impurity vertex $W(\mathbf{p},\mathbf{p}',\mathbf{q})$ [Eq. (19)]; (b) diagram of lowest order for the self-energy part of the excitonic propagator. The dashed line correspond to the correlation function $W(\mathbf{p},\mathbf{p}',0)$.

and

$$\alpha_{e(h)} = -\frac{2m_{e(h)}}{M}\kappa, \quad \kappa = \frac{\hbar^2}{Ma_B^4}\sum_n{}' \frac{|\langle 0|x|n\rangle|^2}{\epsilon_0-\epsilon_n}. \quad (15)$$

Here $n$ denotes the exciton excited states. Exact calculations of the dimensionless constants $\alpha$ and $\beta$ for a 2D Wannier–Mott exciton are given in Appendix B. Note that $\beta_e$, $\beta_h > 0$ are positive; therefore, exciton scattering increases with $B$ when $\ell_B \gg a_B$.

Using perturbation theory, one can also obtain the exciton density of states in a magnetic field. The exciton spectrum is given by the formula

$$\epsilon(p) = -\epsilon_0\left[1-\left(\frac{l_2}{\ell_B}\right)^4\right] + \frac{\hbar^2 p^2}{2M}\left[1-\kappa\left(\frac{a_B}{\ell_B}\right)^4\right], \quad (16)$$

where the parameter $l_2 = 3a_B/8$ determines the diamagnetic shift. The 2D-exciton density of states derives from the second term on the right-hand side of Eq. (16):

$$\mathcal{N}(\epsilon) = \frac{2M/\hbar^2}{1-\kappa(a_B/\ell_B)^4}. \quad (17)$$

The exciton mass and, hence, the density of states $\mathcal{N}(\epsilon)$, increase with the magnetic field $B$. As will be shown below, it is this tendency that generally determines the change in the classical diffusion constant in weak magnetic fields.

### 2.3. Cooperon: weak-field limit

The approximation in which the complete vertex $\Gamma$ (Fig. 1) is replaced by a sum of ladder diagrams (a diffuson) corresponds to the description of transport based on the Boltzmann equation (see, e.g., Refs. 23 and 26). This approximation yields the "classical" diffusion constant, which does not take into account the interference of different paths. If the random potential is weak, all other diagrams with crossed impurity lines have smallness[3] of order $\gamma/\epsilon \ll 1$. The only exception is the class of maximally crossed diagrams in the electron-hole channel,[2] which determines quantum weak-localization corrections to the diffusion constant. The complete sum of such diagrams (the cooperon) is shown in Fig. 2. The exceptional role of these diagrams is due to the following fact: when the total momentum $\mathbf{p}+\mathbf{p}'-\mathbf{q} \approx 0$, the Green's functions $G^R$ and $G^A$ for the maximally crossed diagrams are always grouped in pairs with close poles by virtue of momentum conservation. As a result, they provide a "resonant" contribution after integration.

The damping coefficient for an exciton with a momentum $\mathbf{p}$ is determined by the imaginary part of the self-energy part (Fig. 3b):

$$\gamma_0(p) = -\mathrm{Im}\int \frac{d\mathbf{p}'}{(2\pi)^2}\frac{W(\mathbf{p},\mathbf{p}',0)}{\epsilon-\epsilon(p')+i\gamma_0(p')}. \quad (18)$$

Here $\epsilon(p)$ is the dispersion law (16), and $W(\mathbf{p},\mathbf{p}_1,\mathbf{q})$ is the correlation function of the scattering potential (Fig. 3a):

$$W(\mathbf{p},\mathbf{p}_1,\mathbf{q}) \equiv \langle\langle V_{\mathbf{p},\mathbf{p}_1} V_{\mathbf{p}_1-\mathbf{q},\mathbf{p}-\mathbf{q}}\rangle\rangle. \quad (19)$$

In the weak-field limit discussed in this paper, it has the form[4)]

$$W(\mathbf{p},\mathbf{p}',0) = B_{ee}(\Delta\mathbf{p})\left[1+\left(\frac{a_B}{\ell_B}\right)^4(2\beta_e(\Delta\mathbf{p})^2 a_B^2+\alpha_e^2\right.$$
$$\left.\times(\mathbf{p}\times\mathbf{p}')_z^2 a_B^4\right] + B_{hh}(\Delta\mathbf{p})\left[1+\left(\frac{a_B}{\ell_B}\right)^4\right.$$
$$\left.\times(2\beta_h(\Delta\mathbf{p})^2 a_B^2+\alpha_h^2(\mathbf{p}\times\mathbf{p}')_z^2 a_B^4)\right] + B_{eh}(\Delta\mathbf{p})$$
$$\times\left[1+\left(\frac{a_B}{\ell_B}\right)^4((\beta_e+\beta_h)(\Delta\mathbf{p})^2 a_B^2 -\alpha_e\alpha_h\right.$$
$$\left.\times(\mathbf{p}\times\mathbf{p}')_z^2 a_B^4\right] + B_{he}(\Delta\mathbf{p})\left[1+\left(\frac{a_B}{\ell_B}\right)^4\right.$$
$$\left.\times((\beta_e+\beta_h)(\Delta\mathbf{p})^2 a_B^2-\alpha_e\alpha_h(\mathbf{p}\times\mathbf{p}')_z^2 a_B^4)\right], \quad (20)$$

where $B_{ij}(\mathbf{p}) = \langle\langle\bar{V}_i(\mathbf{p})\bar{V}_j(-\mathbf{p})\rangle\rangle$. As usual, if $\gamma_0 \ll \epsilon$, we have

$$\gamma_0(p) = \pi\mathcal{N}(\epsilon)\int\frac{d\phi_{p_1}}{2\pi}W(\mathbf{p},\mathbf{p}_1,0), \quad (21)$$

where $|\mathbf{p}_1|$ lies on the mass surface $\epsilon(p_1) = \epsilon$, so that only averaging over angles remains in Eq. (18). The effect of the magnetic field $B$ on the damping coefficient $\gamma_0(p)$ can be approximately estimated as follows (we assume that all correlators of random fields $B_{ij}$ are comparable):

$$\gamma_0(p) \approx \gamma_0\frac{1+4(\beta_e+\beta_h)(pa_B)^2(a_B/\ell_B)^4}{1-\kappa(a_B/\ell_B)^4}, \quad (22)$$

where $\gamma_0$ is the damping coefficient in a zero magnetic field. The numerator on the right-hand side of Eq. (22) contains the additional small parameter $(pa_B)^2 \ll 1$ in comparison with the denominator. This means that the main effect of the magnetic field $B$ is due to the growing exciton density of states (increase in exciton mass) with increasing $B$ [see Eq. (16)], whereas the changes in the scattering matrix elements play a minor role.

In the weak-field limit, as in the case of strong magnetic fields,[20] the diffusion pole in the cooperon is absent owing to the broken time-reversal symmetry for the effective potential (12). Let us prove this statement. As usual, it is convenient to write the equation for the cooperon $U$ in variables $\mathbf{p}$, $\mathbf{p}'$, and $\mathbf{K} = \mathbf{p} + \mathbf{p}' - \mathbf{q}$, where $\mathbf{K}$ is the total (conserved) momentum, and $\mathbf{q}$ is the momentum corresponding to density fluctuations. For $U$ we obtain the Bethe–Salpeter equation in the usual manner:

$$U_{\epsilon,\omega}(\mathbf{p},\mathbf{p}',\mathbf{K}) = U^0_{\epsilon,\omega}(\mathbf{p},\mathbf{p}',\mathbf{K})$$
$$+ \int \frac{d\mathbf{p}_1}{(2\pi)^2} \widetilde{W}(\mathbf{p},\mathbf{p}_1,\mathbf{K}) G^R_\epsilon(\mathbf{p}_1) G^A_{\epsilon-\omega}$$
$$\times (\mathbf{K}-\mathbf{p}_1) U_{\epsilon,\omega}(\mathbf{p}_1,\mathbf{p}',\mathbf{K}), \quad (23)$$

where

$$U^0_{\epsilon,\omega}(\mathbf{p},\mathbf{p}',\mathbf{K}) = \int \frac{d\mathbf{p}_1}{(2\pi)^2} \widetilde{W}(\mathbf{p},\mathbf{p}_1,\mathbf{K}) G^R_\epsilon(\mathbf{p}_1) G^A_{\epsilon-\omega}$$
$$\times (\mathbf{K}-\mathbf{p}_1) \widetilde{W}(\mathbf{p}_1,\mathbf{p}',\mathbf{K}) \quad (24)$$

and we have introduced the correlation function:

$$\widetilde{W}(\mathbf{p},\mathbf{p}_1,\mathbf{K}) \equiv \langle\langle V_{\mathbf{p},\mathbf{p}_1} V_{\mathbf{K}-\mathbf{p},\mathbf{K}-\mathbf{p}_1}\rangle\rangle. \quad (25)$$

In contrast to the conventional theory, the system is characterized by two correlation functions $W$ [Eq. (20)] and $\widetilde{W}$.[5] The difference between the correlation functions is caused by the broken time-reversal symmetry for the effective scattering potential (13). As a result, the terms with the vector product $(\mathbf{p}\times\mathbf{p}_1)_z$ in $\widetilde{W}(\mathbf{p},\mathbf{p}_1,\mathbf{K}=0)$ have signs opposite to those of the terms in $W(\mathbf{p},\mathbf{p}_1,\mathbf{q}=0)$.

In the limit of weak disorder, we have $G^R G^A \sim \delta(\epsilon(p) - \epsilon)$, so that the integration in Eq. (23) is reduced to averaging over angles. In the usual case the isotropic (with respect to $\mathbf{p},\mathbf{p}'$) part of $U_{\epsilon,\omega}(\mathbf{p},\mathbf{p}_1,\mathbf{K})$ diverges as $\mathbf{K},\omega\to 0$. This happens because the following relation holds:

$$\int \frac{d\mathbf{p}_1}{(2\pi)^2} W(\mathbf{p},\mathbf{p}_1,0) G^R_\epsilon(\mathbf{p}_1) G^A_\epsilon(-\mathbf{p}_1) = 1. \quad (26)$$

Then it follows from Eqs. (23) and (26) (if $\widetilde{W} = W$) that $\int d\phi_\mathbf{p} \int d\phi_{\mathbf{p}_1} U_{\epsilon,\omega}(\mathbf{p},\mathbf{p}_1,0) \to \infty$. In the case under consideration, however, the isotropic part of $U$ is finite in the limit $\mathbf{K},\omega\to 0$. In fact, using the identity $\widetilde{W} \equiv W + (\widetilde{W} - W)$, we obtain

$$\int \frac{d\mathbf{p}_1}{(2\pi)^2} \widetilde{W}(\mathbf{p},\mathbf{p}_1,0) G^R_\epsilon(\mathbf{p}_1) G^A_\epsilon(-\mathbf{p}_1) = 1 - \frac{\gamma_B}{\gamma_0}, \quad (27)$$

where $\gamma_B(p) = \gamma_0(p) - \widetilde{\gamma}_0(p) \geq 0$,

$$\widetilde{\gamma}_0(p) = \pi \mathcal{N}(\epsilon) \int \frac{d\phi_{\mathbf{p}_1}}{2\pi} \widetilde{W}(\mathbf{p},\mathbf{p}_1,0). \quad (28)$$

If, however, $\gamma_B \ll \gamma_0$, the isotropic part of $U$ still makes the principal contribution. A solution of Eq. (23) in the region of low frequencies $\omega$ and small momenta $\mathbf{K}$, which is discussed in the paper, can be obtained using an expansion in terms of angular momenta, and was described in detail in Ref. 20 (see also Ref. 27). Ultimately, the expression for the cooperon has the form

$$U(\mathbf{K},\omega) = \frac{2\widetilde{\gamma}_0 \gamma_0/\pi\mathcal{N}(\epsilon)}{D^c K^2 - i\omega + 2\gamma_B \gamma_0/\widetilde{\gamma}_0}. \quad (29)$$

Here

$$D^c = p^2/4M^2 \widetilde{\gamma}_{\text{tr}}, \quad \widetilde{\gamma}_{\text{tr}} = \gamma_0 - \widetilde{\gamma}_1 \geq 0, \quad (30)$$

$$\widetilde{\gamma}_1 = 2 \int \frac{d\phi_\mathbf{p}}{2\pi} \int \frac{d\phi_{\mathbf{p}_1}}{2\pi} (\hat{\mathbf{p}}\hat{\mathbf{K}}) \widetilde{W}(\mathbf{p},\mathbf{p}_1,0)(\hat{\mathbf{p}}_1 \hat{\mathbf{K}}), \quad (31)$$

where $\hat{\mathbf{p}} = \mathbf{p}/|\mathbf{p}|$. The special feature of this solution is that it contains a finite dephasing time $\gamma_B^{-1}$ for a neutral composite particle in a magnetic field $B$, and this dephasing time eliminates a singularity, namely the diffusion pole. Formally, this case is similar to that of electron scattering by magnetic impurities.[8,9,11,12]

In weak magnetic fields, $\gamma_B$ can be estimated using explicit expressions for $W$ and $\widetilde{W}$:

$$\gamma_B(p) \simeq (pa_B)^4 \left(\frac{a_B}{\ell_B}\right)^4 \gamma_0(p). \quad (32)$$

The emergence of the characteristic dephasing time $\tau_B = \hbar/\gamma_B$ estimated by Eq. (32) can be interpreted in qualitative terms as follows. An exciton acquires a random phase in a magnetic field only as a result of impurity scattering. When an exciton with momentum $\mathbf{p}$ is scattered by an impurity, its kinetic energy $E_{\text{kin}} = \hbar^2 p^2/2M$ can be treated as a perturbation to the internal electron-hole motion with an energy $E_{\text{exc}} = \epsilon_0$. This results in fluctuations in the mean square distance between $e$ and $h$: $\langle\langle\Delta r^2\rangle\rangle \sim (E_{\text{kin}}/E_{\text{exc}})a_B^2$. This additional separation between the electron and hole orbits due to a scattering act leads to an increase in the magnetic flux passing "through" the exciton $\Delta\Phi \sim \langle\langle\Delta r^2\rangle\rangle B$, which corresponds to a (random) phase shift $\Delta\phi \sim \langle\langle\Delta r^2\rangle\rangle B/\Phi_0$ in the wave function, where $\Phi_0$ is the magnetic flux quantum. Therefore, the random phase shift in a single scattering act, which takes place during the time interval $\tau = \hbar/\gamma$, is $\sim (\Delta\Phi/\Phi_0) = (pa_B)^2(a_B/\ell_B)^2 \ll 1$. Since the phase shifts of the wave function are random, the total phase shift becomes comparable to unity and coherence is lost only after $(\Phi_0/\Delta\Phi)^2$ scattering acts. The corresponding characteristic time $\tau_B \sim (\Phi_0/\Delta\Phi)^2 \tau \sim \tau(pa_B)^{-4}(\ell_B/a_B)^4$, which is consistent with Eq. (32).

### 2.4. Diffusion constant

In order to obtain quantum corrections to the diffusion constant, one should include Eq. (29) for the cooperon $U(\mathbf{K},\omega)$ together with the first-order impurity vertex $W(\mathbf{p},\mathbf{p}',\mathbf{K})$ in the ladder diagrams (Figs. 1 and 4) for the effective conductivity.[3] In the ladder approximation, we have a transport coefficient $\gamma_{tr}$ instead of $\gamma$ for a random field with a finite correlation length. Technical details of the diagram treatment for 2D excitons are given in Appendix C. The diffusion constant for an exciton of energy $\epsilon$ takes the form

$$D(\epsilon) = D_0(\epsilon)\left[1 + \frac{\tilde{\gamma}_0}{4\pi^2 \gamma_{tr} \mathcal{N}(\epsilon) D^c} \ln\left(\frac{D^c K_0^2 \tilde{\gamma}_0}{2\gamma_B \gamma_0}\right)\right]^{-1},\quad (33)$$

where $K_0 \simeq \gamma(p)/V(p)$ is the cut-off momentum and $D_0 = p^2/4M^2\gamma_{tr}$ is the conventional (''classical'') diffusion constant for an exciton.[19,21]

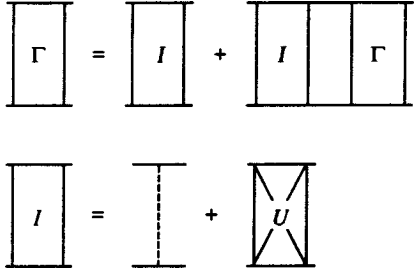

FIG. 4. Diagrammatic representation of the approximation for the vertex Γ including quantum corrections to the diffusion constant.

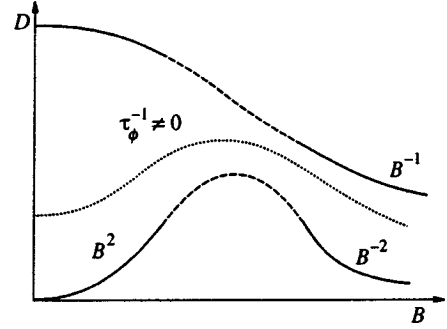

FIG. 5. Static diffusion constant $D$ of excitons as a function of the field $B$ classical transport (upper curve), with consideration of quantum corrections (lower curve), and with weak inelastic scattering (middle curve).

Before discussing the quantum corrections (33), let us derive the classical diffusion constant $D_0$ as a function of the magnetic field $B$. Using general expressions (22) and (16), we obtain

$$D_0(\epsilon,B) \simeq D_0(\epsilon)\left[1 - 3\kappa\left(\frac{a_B}{\ell_B}\right)^4\right] \equiv D_0\left[1 - \left(\frac{B}{B_0}\right)^2\right],\quad (34)$$

where the characteristic magnetic field $B_0$ is determined by the expression $B_0 a_B^2 \simeq \Phi_0$ and $D_0$ is the diffusion constant at $B=0$.[19] The diffusion constant $D_0$ monotonically decreases as the magnetic field increases in accordance with Eq. (34).

The inclusion of quantum corrections drastically changes the dependence of $D$ on $B$. Indeed, $\gamma_B$ tends to zero as $B \to 0$, and $D$ vanishes as a result [see Eq. (33)]; this is the weak localization of excitons in the absence of $B$ (excitons, like ordinary 2D particles, are localized in a random potential[7]). A self-consistent approach may be used in this situation.[27] In fact, the approximation for the complete vertex Γ including only ladder diagrams for $\Gamma_0$ and maximally crossed diagrams for $U$ applies only to the case of weak scattering, where the resulting diffusion constant is large. When the complete vertex corresponds to strong scattering and the diffusion constant $D$ is small, one cannot, strictly speaking, select a preferential class of diagrams. The underlying idea of the self-consistent approach[27] is the existence of a relation between $\Gamma_0$ and $U$ in the presence of time-reversal symmetry (maximally crossed diagrams in the $e-h$ channel are ladder diagrams in the $e-e$ channel). One consequence of this relation is that the diffusion pole, which exists in the diffuson at small momentum transfers, is ''transmitted'' to the cooperon (where it exists at small total momenta $\mathbf{K}$). Since the vertex Γ is directly related to the ''density-density'' correlation function, it is physically clear that it is the diffusion constant $D$ that should appear in Γ, and the constant $D_0$ in the cooperon should consequently also be replaced by $D$. The mathematical basis of this approach was discussed in detail by Suslov.[28]

In the specific case under consideration, time-reversal symmetry is broken, and, strictly speaking, there is no duality between the diffuson and cooperon. We can use, however, the self-consistent approximation in order to obtain the leading terms in the $B$ expansion of the total diffusion constant. The point is that $D_0$ and $D^c$ behave similarly in the leading orders in $B$. Therefore, the diffusion constant $D^c$ in the cooperon can be replaced by the total diffusion constant $D$ so that $D(B)$ could be calculated in a self-consistent manner using Eq. (33). In the case $D(B) \ll D_0$ the magnetic-field dependence is given by

$$D(\epsilon,B) = (pa_B)^4\left(\frac{B}{B_0}\right)^2 D_0(\epsilon)\exp[\mathcal{N}(\epsilon)D_0(\epsilon)],\quad (35)$$

where $B_0 a_B^2 \simeq \Phi_0$ [see Eq. (34)]. Thus, the static diffusion constant at $B=0$ is zero and increases proportionally to $B^2$ at small $B$. This behavior of $D$ corresponds to the suppression of the weak localization of excitons in a magnetic field.

In the strong-field limit, $\ell_B \ll a_B$, the exciton diffusion constant $D$ drops as $B^{-2}$.[20] Thus, it is clear that, if weak-localization effects are taken into account, $D$ is a nonmonotonic function of the magnetic field. Note that the classical diffusion constant $D_0$ decreases monotonically with the magnetic field in both the strong-field limit, $\ell_B \ll a_B$, and in the weak-field limit, $\ell_B \gg a_B$. These results are illustrated by Fig. 5. Weak inelastic processes characterized by the dephasing time $\tau_\phi = \hbar/\gamma_\phi$ can be included in our scheme phenomenologically. If $\tau_\phi$ is finite, the diffusion constant is finite even at $B=0$. The value $D(B=0)$ is controlled by $\gamma_\phi$, which should be added to $\gamma_B$ in Eq. (33). The appearance of $\gamma_\phi$ affects $D_0$ and $D$ differently. If the condition $\gamma_\phi \ll \gamma_0$ holds, this addition has little effect on $D_0$. But, since $D_0 \mathcal{N} \gg 1$ in the weak-scattering limit, the relation $D_0 \mathcal{N} \gamma_\phi \gg \gamma_0$ can hold even when $\gamma_\phi \ll \gamma_0$. In this case, the weak-localization corrections are small, and we have $D(B) \simeq D_0(B)$.

Analytical calculations are impossible in intermediate fields, where $\ell_B \sim a_B$. It is quite natural to assume that in this range the magnetic field dependence of the diffusion constant (either the classical constant, $D_0$, or the constant which takes into account the quantum corrections, $D$) has the form shown by the dashed lines in Fig. 5. The increase in $D$ with the magnetic field $B$ (*a positive magnetodiffusion effect*) is due to the suppression of the weak localization of excitons in magnetic fields. This effect is similar to the negative magnetoresistance in 2D electron systems.[8]

### 3. CONCLUSIONS

We have shown that a magnetic field $B$ eliminates divergence of the maximally crossed diagrams in the "exciton-antiexciton" channel (the exciton analogue of the cooperon). Unlike charged particles, an exciton acquires a phase in the field $B$ not during free motion, but only upon scattering by defects. As a result, the diffusion constant of 2D excitons in magnetic fields remains finite as $\omega \to 0$ (under the assumption that the random potential is weak). The static diffusion constant $D(B)$ is a decreasing function of $B$ in strong magnetic fields, $\ell_B \ll a_B$, whereas in weak magnetic fields, $\ell_B \gg a_B$ (and, probably, in intermediate fields, $\ell_B \sim a_B$) $D(B)$ increases with the magnetic field, i.e., a positive magnetodiffusion effect takes place for excitons. The self-consistent approximation yields $D \propto B^2$ in weak magnetic fields, which indicates that weak localization is suppressed in a magnetic field $B$. Quantum corrections are also important in the strong-field limit, $\ell_B \ll a_B$, and lead to a faster power-law decrease in the diffusion constant with the magnetic field, $D \propto B^{-2}$,[20] in comparison to the classical diffusion constant, $D_0 \propto B^{-1}$.[19,21] This is because the characteristic internal length scale of the magnetoexciton $\ell_B \ll a_B$ decreases with increasing $B$ as $\ell_B \propto B^{-1/2}$, and its internal structure has a lesser impact on the scattering process, so that the magnetoexciton becomes similar to a structureless neutral boson. Thus, for neutral $e-h$ systems, crossover to the exciton weak-localization regime takes place in the strong-field limit (unlike electron systems, which contain delocalized states in the quantum Hall effect regime).

Although the calculated function $D(B)$ is a nonmonotonic function of $B$, it does not reproduce all the details of the experimental findings for $D(B)$.[17] For instance, the observed suppression of exciton magnetotransport in the range of relatively low fields[17] is in agreement with the theoretical predictions for the behavior of the *classical* diffusion constant (Fig. 5). Our calculations, however, demonstrate that the increase in $D(B)$ in the range $B > 6$ T observed in Ref. 17 cannot be interpreted in terms of the suppression of weak localization of excitons in a magnetic field. Note that the localization regime in double quantum wells used in experiments[16–18] is closer to the strong-localization regime of excitons.[19] Also, we have not considered the effects of the Bose–Einstein condensation of excitons. The investigation of the effects of a magnetic field on the strong localization of excitons and of Bose–Einstein condensation on the transport of neutral composite particles (excitons) is a very interesting problem, which has not yet been solved. Note also that, as in the case of electrons in quasi-two-dimensional semiconductor structures,[12] the effects of fast transitions between different spin states may be important for excitons.

Our theoretical prediction of an increase in exciton mobility with increasing $B$ in the weak-localization regime can be tested experimentally at low temperatures (where inelastic scattering is suppressed and the dephasing time $\tau_\phi$ is large) in magnetic fields for which $\ell_B \gtrsim a_B$. Such experiments require quantum wells with a weak random potential, for example, wide quantum wells with smooth interfaces.

We are indebted to G. E. W. Bauer, L. V. Butov, E. L. Ivchenko, Yu. V. Nazarov, and S. G. Tikhodeev for useful discussions. This work was supported by Volkswagen Stiftung (Grant VW I/69 361), the Nederlandse Organisatie voor Wetenschappelijk Onderzoek (Netherlands Organization for Scientific Research) (Grant NWO 047-003-018), INTAS-RBRF (Grant 95-675), and the Russian Fund for Fundamental Research.

### APPENDIX A: DIAGRAMMATIC REPRESENTATION OF EXCITON SCATTERING

To the best of our knowledge, no approximation which reduces the scattering of excitons to an effectively one-particle process has been rigorously developed using diagram techniques. An approach similar to that developed in this Appendix can be applied to other problems, such as investigations of the role of transitions to excited states, the effects of a finite exciton density, and strong localization in terms of the effective exciton scattering.

In the electron-hole representation, a Wannier–Mott exciton is described in terms of a sum of ladder diagrams, which include the $e-h$ Coulomb interaction. The corresponding two-particle Green's function can be expanded in terms of the exciton eigenfunctions $\Psi_\lambda(\mathbf{r}_e, \mathbf{r}_h)$:

$$G_2(\mathbf{r}_1,\mathbf{r}_2,t;\mathbf{r}_3,\mathbf{r}_4,t')$$
$$\equiv -i\langle T\hat{\Psi}_e(\mathbf{r}_1,t)\hat{\Psi}_h(\mathbf{r}_2,t)\hat{\Psi}_h^\dagger(\mathbf{r}_4,t')\hat{\Psi}_e^\dagger(\mathbf{r}_3,t')\rangle$$
$$= \int dE \sum_\lambda \frac{\Psi_\lambda^*(\mathbf{r}_1,\mathbf{r}_2)\Psi_\lambda(\mathbf{r}_3,\mathbf{r}_4)}{E-\epsilon_\lambda}\exp[-iE(t-t')], \tag{A1}$$

where $\hat{\Psi}_e^\dagger(\mathbf{r},t)$ and $\hat{\Psi}_h^\dagger(\mathbf{r},t)$ are the electron and hole creation operators in the Heisenberg representation. For simplicity, we consider the case of zero magnetic field, $B=0$, where $\epsilon_\lambda$ are the ordinary eigenvalues of the exciton energies, $\epsilon_\lambda = \epsilon(p) + \epsilon_n$, $\epsilon_n$ is the energy (in either the discrete or continuous spectrum) of relative motion, $\epsilon(p)$ is the center-of-mass kinetic energy, $\Psi_\lambda(\mathbf{r}_1, \mathbf{r}_2) = \exp(i\mathbf{p}\mathbf{R})\Phi_n(\mathbf{r})$ are the exciton wave functions, $\mathbf{R} = (m_e\mathbf{r}_1 + m_h\mathbf{r}_2)/M$, and $\mathbf{r} = \mathbf{r}_1 - \mathbf{r}_2$. Our aim is to replace the two-particle $e-h$ Green's function by an effective "one-particle" Green's function of an exciton defined by the formula

$$G(\mathbf{R},t;\mathbf{R}',t') = -i\langle TB(\mathbf{R},t)B^\dagger(\mathbf{R}',t')\rangle, \tag{A2}$$

where the exciton creation operator is defined as

$$B_\lambda^\dagger(\mathbf{R},t) = \int d\mathbf{r}\,\hat{\Psi}_e^\dagger\!\left(\mathbf{R} + \frac{m_h}{M}\mathbf{r}, t\right)$$
$$\times \hat{\Psi}_h^\dagger\!\left(\mathbf{R} - \frac{m_e}{M}\mathbf{r}, t\right) \Psi_\lambda^*(\mathbf{R},\mathbf{r}). \tag{A3}$$

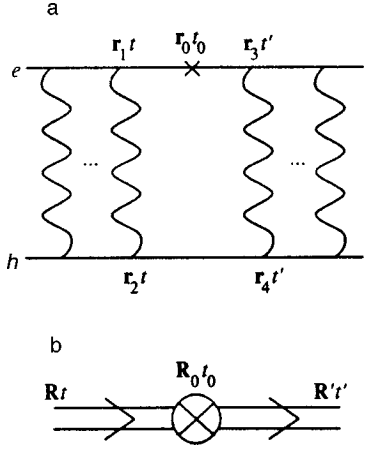

FIG. 6. (a) Impurity vertex in the electron line of the two-particle $e$–$h$ propagator. The wavy lines correspond to the Coulomb $e$–$h$ interaction. (b) Impurity vertex (10) in the excitonic propagator.

The function $G_2$ satisfies the following Bethe–Salpeter equation:

$$G_2(\mathbf{r}_1,\mathbf{r}_2,t;\mathbf{r}_3,\mathbf{r}_4,t')$$
$$= G_e(\mathbf{r}_1,\mathbf{r}_3,t-t')G_h(\mathbf{r}_2,\mathbf{r}_4,t-t')$$
$$+ \int d\mathbf{r}_3'd\mathbf{r}_4'dt_1' G_e(\mathbf{r}_1,\mathbf{r}_3',t-t_1')G_h(\mathbf{r}_2,\mathbf{r}_4',t-t_1')$$
$$\times U(\mathbf{r}_3'-\mathbf{r}_4')G_2(\mathbf{r}_3',\mathbf{r}_4',t_1';\mathbf{r}_3,\mathbf{r}_4,t'). \tag{A4}$$

Applying the operator (A4)

$$[G_e(\mathbf{r}_1,t)G_h(\mathbf{r}_2,t)]^{-1} = i\frac{\partial}{\partial t} + \frac{\nabla_1^2}{2m_e} + \frac{\nabla_2^2}{2m_h}$$

to both sides, we obtain the Schrödinger-like equation:

$$\left[i\frac{\partial}{\partial t} + \frac{\nabla_1^2}{2m_e} + \frac{\nabla_2^2}{2m_h} - U(\mathbf{r}_1-\mathbf{r}_2)\right]G_2(\mathbf{r}_1,\mathbf{r}_2,t;\mathbf{r}_3,\mathbf{r}_4,t')$$
$$= \delta(\mathbf{r}_1-\mathbf{r}_3)\delta(\mathbf{r}_2-\mathbf{r}_4)\delta(t-t'). \tag{A5}$$

Now let us consider a diagram with only one impurity vertex corresponding to the external potential $V_e$ in the electron line (Fig. 6a).

The set of external coordinates $(\mathbf{r}_e,\mathbf{r}_h,t)$ will be symbolically denoted by $\bar{X},\bar{X}'$. The analytical expression for $\widetilde{G}_2(\bar{X},\bar{X}')$ in the case of the diagram in Fig. 6a has the form

$$\widetilde{G}_2(\bar{X},\bar{X}') = \int d\mathbf{r}_0 d\mathbf{r}_1 d\mathbf{r}_2 d\mathbf{r}_3 d\mathbf{r}_4 dt dt_0 dt' G_2(\bar{X};\mathbf{r}_1,\mathbf{r}_2,t)$$
$$\times U(\mathbf{r}_1-\mathbf{r}_2)G_e(\mathbf{r}_1-\mathbf{r}_0,t-t_0)V_e(\mathbf{r}_0)$$
$$\times G_e(\mathbf{r}_0-\mathbf{r}_3,t_0-t')G_h(\mathbf{r}_2-\mathbf{r}_4,t-t')$$
$$\times U(\mathbf{r}_3-\mathbf{r}_4)G_2(\mathbf{r}_3,\mathbf{r}_4,t';\bar{X}'). \tag{A6}$$

Using Eq. (A5), we can replace the function $G_2U$ by a differential operator acting on $G_e$ and $G_h$. Given that

$$\left(i\frac{\partial}{\partial t} + \frac{\nabla^2}{2m_{e(h)}}\right)G_{e(h)}(\mathbf{r}-\mathbf{r}',t-t') = \delta(t-t')\delta(\mathbf{r}-\mathbf{r}'),$$

we obtain the following expression for function (A6):

$$\widetilde{G}_2(\bar{X},\bar{X}') = \int d\mathbf{r}_1 d\mathbf{r}_2 d\mathbf{r}_3 d\mathbf{r}_4 dt\,dt' G_2(\bar{X};\mathbf{r}_1,\mathbf{r}_2,t)$$
$$\times V_e(\mathbf{r}_1)G_e(\mathbf{r}_1-\mathbf{r}_3,t-t')G_h(\mathbf{r}_2-\mathbf{r}_4,t-t')$$
$$\times U(\mathbf{r}_3-\mathbf{r}_4)G_2(\mathbf{r}_3,\mathbf{r}_4,t';\bar{X}') + \int d\mathbf{r}_0 d\mathbf{r}_1 d\mathbf{r}_2$$
$$\times d\mathbf{r}_3 dt\,dt_0 G_2(\bar{X};\mathbf{r}_1,\mathbf{r}_2,t)G_e(\mathbf{r}_1-\mathbf{r}_0,t-t_0)$$
$$\times V_e(\mathbf{r}_0)G_e(\mathbf{r}_0-\mathbf{r}_3,t_0-t)U(\mathbf{r}_3-\mathbf{r}_2)$$
$$\times G_2(\mathbf{r}_3,\mathbf{r}_2,t;\bar{X}'). \tag{A7}$$

Note that the second term contains the product $G_e(t) \times G_e(-t)$, which contributes a factor $\sim n_e(1-n_e)$ and can, therefore, be neglected in the low-density limit. Thus only the first term remains on the right-hand side of Eq. (A7). Let us also take into account that the expansion for $G_2$ begins with a term of zero order in $G_eG_h$ [corresponding to $\delta$ functions on the right-hand side of Eq. (A5)] and add it to Eqs. (A6) and (A7). Then we can see that the first term of the right-hand side of Eq. (A7) contains the Coulomb ladder diagrams on both sides of the impurity vertex $V_e(\mathbf{r}_1)$. In addition, the temporal and spatial coordinates coincide in such a manner that, using representations (A2) and (A5), we can represent expression (A3) in the form of a diagram corresponding to scattering of an exciton as a whole (Fig. 6b). After adding the analogous term for scattering of the hole, we can see that the effective scattering potential in the excitonic representation is indeed determined by Eqs. (9) and (10).

## APPENDIX B: CALCULATION OF PERTURBATION SERIES

In a perturbative analysis of systems with 2D excitons in a magnetic field, sums like those in Eqs. (14) and (15) appear frequently. Therefore, it seems useful to calculate these sums exactly for the case of a two-dimensional hydrogenic exciton. If the operator approach[29] is applied, the explicit form of the ground-state wave function is sufficient. In the intermediate calculations we set $a_B = \hbar = 1$ and return to dimensional quantities in the final expressions. Let us start with the constant $\kappa$ in Eq. (15). If we can find the explicit form of the operator $\hat{b}$ that satisfies the quantum equation of motion $\mu\partial\hat{b}/\partial t = i\mu[H_0,\hat{b}] = x$, where $H_0$ is the Hamiltonian of a 2D hydrogen atom (7) in the absence of a field $B$, for the matrix elements we have

$$i\mu(\epsilon_0 - \epsilon_n)\langle 0|\hat{b}|n\rangle = \langle 0|x|n\rangle, \tag{B1}$$

[where $\epsilon_n = \epsilon_0/(n+1/2)^2$ and $\epsilon_0 = -\mu e^4/2\varepsilon^2\hbar^2$], and the sum in Eq. (15) is reduced to the diagonal matrix element

$$\kappa = i\frac{\mu}{M}\sum_n{}' \langle 0|x|n\rangle\langle n|\hat{b}|0\rangle = i\frac{\mu}{M}\langle 0|x\hat{b}|0\rangle, \quad (B2)$$

where the prime means that state $n=0$ is not included in the summation. We have used the completeness condition for the states $\Sigma_n |n\rangle\langle n| = 1$ and the equality $\langle 0|x|0\rangle = 0$. In the coordinate representation, we introduce the notation $\hat{b}\phi_0(r) \equiv b(\mathbf{r})\phi_0(r)$, where $\phi_0(r) = \sqrt{8/\pi}\exp(-2r)$ is the ground-state wave function. Using the explicit form of the Hamiltonian, we obtain a differential equation for $b(\mathbf{r})$:

$$\frac{1}{2}[\nabla^2 b(\mathbf{r})]\phi_0(r) + (\nabla b(\mathbf{r})\cdot\nabla\phi_0(r)) = -ix\phi_0(r), \quad (B3)$$

whence it follows that $b(\mathbf{r}) = -ib(r)\cos\phi$, and the unknown function $b(r)$ is given by the equation

$$b''(r) + b'(r)\left(\frac{1}{r}-4\right) - \frac{b(r)}{r^2} - 2r = 0. \quad (B4)$$

Solving Eq. (B4), we obtain

$$b(\mathbf{r}) = -i\left(\frac{1}{4}r^2 + \frac{3}{16}r\right)\cos\phi. \quad (B5)$$

Finally, the matrix element in Eq. (B2) is expressed as

$$\kappa = \frac{\mu}{M}\int_0^\infty dr\left(\int_0^{2\pi}d\phi\frac{8}{\pi}\cos^2\phi\right)$$
$$\times\exp(-4r)\left(\frac{1}{4}r^2 + \frac{3}{16}r\right)r^2 = \frac{21}{16^2}\frac{\mu}{M}, \quad (B6)$$

and the coefficient $\alpha$ in Eq. (15) is given by the expression

$$\alpha_{e(h)} = 2\frac{21}{16^2}\frac{m_{e(h)}\mu}{M^2}.$$

The same operator $\hat{b}$ can be used in calculating the first sum in Eq. (14):

$$I_1 = \sum_n{}'\frac{|\langle 0|x|n\rangle|^2}{(\epsilon_0 - \epsilon_n)^2}. \quad (B7)$$

Using Eq. (B1), we can also reduce $I_1$ to a diagonal matrix element: $I_1 = \mu^2\langle 0|\hat{b}\hat{b}|0\rangle$. In combination with the explicit expression (B5) for $b(r)$, this equation yields

$$I_1 = \mu^2\int_0^\infty dr|b(r)|^2\phi_0^2(r) = \frac{159}{4^6}\mu^2. \quad (B8)$$

In order to calculate the second sum in Eq. (14),

$$I_2 = \sum_n{}'\frac{|\langle 0|r^2|n\rangle|^2}{\epsilon_n - \epsilon_0}, \quad (B9)$$

we must find an operator $\hat{b}_2$ such that $i\mu[H_0,\hat{b}_2] = r^2$. We set $\hat{b}_2\phi_0(r) = b_2(r)\phi_0(r)$. Then

$$b_2''(r) + b_2'(r)\left(\frac{1}{r}-4\right) - 2ir^2 = 0.$$

The solution is the function

$$b_2(r) = -\frac{i}{2}\left(\frac{1}{3}r^3 + \frac{3}{8}r^2 + \frac{3}{8}r + \frac{3}{32}\ln r + c_1\right) \quad (B10)$$

with an undetermined constant $c_1$. One feature of calculations of sums like $I_2$ is that, using the completeness condition for the intermediate states, we should eliminate the matrix element of the ground state $n=0$, which does not automatically equal zero, unlike the coordinate matrix element $\langle 0|x|0\rangle = 0$. Therefore, we have [cf. Eq. (B2)]

$$I_2 = i\mu(\langle 0|\hat{b}_2 r^2|0\rangle - \langle 0|\hat{b}_2|0\rangle\langle 0|r^2|0\rangle). \quad (B11)$$

We see that, as a result of subtraction, the final expression does not include the constant $c_1$ introduced in Eq. (B10). This allows us to obtain the exact expression $I_2 = 105\mu/2^9$, and for $\beta$ we have

$$\beta_{e(h)} = \frac{1}{4^6 M^2}\left(105 m_{e(h)}^2 - \frac{159}{2}\mu^2\right). \quad (B12)$$

Note that the coefficients $\beta_{e(h)} > 0$ are always positive [since $\mu = m_e m_h/(m_e + m_h) < m_e, m_h$], but numerically small: $\beta \lesssim 0.02$.

### APPENDIX C: CALCULATION OF THE DIFFUSION CONSTANT $D$

This section gives details of the calculation of the diffusion constant $D$. The calculation of $D(\epsilon)$ should include, in addition to the diagrams of Fig. 4, the zero-order diagram $G^R G^A$. Therefore, the diffusion constant $D(\epsilon)$ is given by the expression

$$D(\epsilon) = \frac{1}{2\pi\mathcal{N}(\epsilon)}\int d\mathbf{p}\int d\mathbf{p}'\frac{\mathbf{p}\mathbf{p}'}{M^2}|G^R(\mathbf{p})|^2[\delta(\mathbf{p}-\mathbf{p}')$$
$$+ \Gamma(\mathbf{p},\mathbf{p}')|G^R(\mathbf{p}')|^2]. \quad (C1)$$

If the cooperon is included in the irreducible part, the complete vertex $\Gamma$ satisfies the Bethe–Salpeter equation shown in Fig. 4. Note that the cooperon (as a function of the variables $\mathbf{p}$ and $\mathbf{p}'$ at $\mathbf{q}=0$) can be expressed approximately as

$$U_{\epsilon,\omega}(\mathbf{p},\mathbf{p}';\mathbf{p}',\mathbf{p}) \simeq \int\frac{d\mathbf{K}}{(2\pi)^2}U(\mathbf{K},\omega)\delta(\mathbf{p}+\mathbf{p}')$$
$$\equiv U\delta(\mathbf{p}+\mathbf{p}'). \quad (C2)$$

This approximation can be used because there are essentially different momentum scales in the problem. Indeed, in integrals like

$$\int d\mathbf{p}\int d\mathbf{p}'|G^R(\mathbf{p})|^2|G^R(\mathbf{p}')|^2 U_{\epsilon,\omega}(\mathbf{p},\mathbf{p}';\mathbf{p}',\mathbf{p})$$
$$= \int d\mathbf{p}\int d\mathbf{K}|G^R(\mathbf{p})|^2|G^R(\mathbf{K}-\mathbf{p})|^2 U_{\epsilon,\omega}(\mathbf{p},\mathbf{K}-\mathbf{p};\mathbf{K}) \quad (C3)$$

only small $\mathbf{K}$ are important owing to the presence of the diffusion pole in $U$. In this case, we can assume in an approximation that $\mathbf{K}-\mathbf{p}\simeq -\mathbf{p}$ and perform integration over $\mathbf{p}$ and **K** independently. This yields Eq. (C2) for the cooperon. Then the equation for the vertex $\Gamma(\mathbf{p},\mathbf{p}')$ shown in Fig. 4 takes the form

$$\Gamma(\mathbf{p},\mathbf{p}') = W(\mathbf{p},\mathbf{p}',0) + U\delta(\mathbf{p}+\mathbf{p}')$$
$$+ \int \frac{d\mathbf{p}_1}{(2\pi)^2}[W(\mathbf{p},\mathbf{p}_1,0) + U\delta$$
$$\times(\mathbf{p}+\mathbf{p}_1)]G^R(\mathbf{p}_1)G^A(\mathbf{p}_1)\Gamma(\mathbf{p}_1,\mathbf{p}'). \quad (C4)$$

The quantity needed for the calculation of $D(\epsilon)$ has the form [see Eq. (C1)]

$$\Gamma_1 = \int \frac{d\phi_{\mathbf{p}}}{2\pi} \int \frac{d\phi_{\mathbf{p}'}}{2\pi}(\hat{\mathbf{p}}\cdot\hat{\mathbf{p}}')\Gamma(\mathbf{p},\mathbf{p}'),$$

where the integration is performed on the mass surface $\epsilon(\mathbf{p}) = \epsilon(\mathbf{p}') = \epsilon$. For the term corresponding to the allowance for the first angular momentum in $\Gamma_1$, Eq. (C4) gives

$$\Gamma_1 = \frac{\gamma_1}{\pi\mathcal{N}_\epsilon} - \frac{U}{2\pi\mathcal{N}_\epsilon\gamma_0} + \frac{\gamma_1}{\gamma_0}\Gamma_1 - \frac{U}{2\gamma_0^2}\Gamma_1, \quad (C5)$$

where

$$\gamma_1 = \int \frac{d\phi_{\mathbf{p}}}{2\pi} \int \frac{d\phi_{\mathbf{p}'}}{2\pi}(\hat{\mathbf{p}}\cdot\hat{\mathbf{p}}')W(\mathbf{p},\mathbf{p}',0).$$

The solution of Eq. (C5) is

$$\Gamma_1 = \frac{\gamma_0}{\pi\mathcal{N}_\epsilon}\frac{\gamma_1 - U/2\gamma_0}{\gamma_{\text{tr}} + U/2\gamma_0}, \quad (C6)$$

where, as usual, $\gamma_{\text{tr}} = \gamma_0 - \gamma_1$. Using Eq. (C6), from Eq. (C1) we obtain

$$D(\epsilon) = D_0\left[1 + \frac{U}{2\gamma_{\text{tr}}\gamma_0}\right]^{-1}. \quad (C7)$$

This allows us to perform the last step of the calculation: by substituting the expressions (C2) and (29) into Eq. (C7), we obtain Eq. (33).

---

*)E-mail: dzyub@gpi.ac.ru

[1] Hereafter we assume that the valence band is nondegenerate and holes have spin 1/2. We ignore the effects due to the different spin states of the exciton, which should be taken into account if the relaxation between them is fast. In the case of III–V semiconductors, this analysis should take account of the real valence band spectrum (see, e.g., Ref. 15), which is beyond the scope of the present study.

[2] When excitons are scattered by charged impurities with a 2D density $n_{\text{imp}}$, we have[19] $\gamma_0/\epsilon \simeq \nu_{\text{imp}}$, where $\nu_{\text{imp}} = 2\pi a_B^2 n_{\text{imp}}$ ($\nu_{\text{imp}} = 2\pi\ell_B^2 n_{\text{imp}}$) is the dimensionless density of impurities for the limiting case $\ell_B \gg a_B$ ($\ell_B \ll a_B$). The smallness of parameter $\nu_{\text{imp}}$ ensures the applicability of the weak-scattering approximation to this scattering mechanism.

[3] In a strong field, $\ell_B \ll a_B$, the term for the Coulomb interaction in Eq. (7) is treated as a perturbation,[19,20] and the wave functions $\Phi_{\mathbf{p}}(\mathbf{r})$ obtained in this limit describe 2D magnetoexcitons.[25]

[4] Equation (20) may include, in principle, terms linear in the magnetic field $B$ [see Eq. (12)]. This is possible, however, only under the peculiar condition that there be a preferential direction in space in the system so that terms linear in $B$ would not vanish in calculating the correlators of $V_{e(h)}$. We do not consider such a possibility in this paper. Note that a preferential direction can be assigned in a system, for example, by an applied electric field.

[5] Otherwise (as in the case of exciton localization in a zero magnetic field[7]), we would simply have the conventional weak-localization theory in an effective potential with a finite correlation length comparable to the exciton radius.

---